\documentclass{aa}  

\usepackage{graphicx}
\usepackage{txfonts}
\usepackage{units}
\usepackage{soul}
\setstcolor{red}
\usepackage{keyval}
\usepackage{natbib}
\usepackage{hyperref} 
\usepackage{xcolor}
\usepackage{xspace}
\newcommand{\mwc}{MWC\,656\xspace}
\newcommand{\fermilat}{\textit{Fermi}-LAT\xspace}
\newcommand{\fermi}{\textit{Fermi}\xspace}
\newcommand{\gammaray}{$\gamma$-ray\xspace}
\newcommand{\kms}{km\,s$^{-1}$\xspace}

\defcitealias{Ribo2017}{R17}
\defcitealias{Munar-Adrover2014}{M14}
\defcitealias{Dzib2015}{D15}

\begin{document}

\title{
  On the nature and Galactic origin of the Be binary \mwc
}
\subtitle{
    New insights from VLA, \textit{Gaia}, and \fermilat
}

\titlerunning{Radio, astrometric, and \gammaray view of \mwc}


\author{
  Sergio A. Dzib\inst{1} \and
  Frederic Jaron\inst{2,1} 
}

\institute{
  Max-Planck-Institut f\"ur Radioastronomie, Auf dem H\"ugel 69, D-53121 Bonn, Germany\\
  \email{sdzib@mpifr-bonn.mpg.de}
  \and
  Department of Geodesy and Geoinformation, Technische Universit\"at Wien (TU Wien), Wiedner Hauptstra\ss e 8-10, 1040, Vienna, Austria\\
  \email{frederic.jaron@tuwien.ac.at}
}

\date{Received ; accepted }


\abstract
{
The binary star \mwc was initially proposed as the first confirmed system composed of a Be star and a black hole. However, recent studies have challenged this interpretation, suggesting that the compact companion is unlikely to be a black hole. In this study, we revisit the nature of \mwc by analyzing archival data across multiple wavelengths, including radio observations from the VLA, optical astrometry from the \textit{Gaia} satellite, and high-energy \gammaray data from the \fermilat.
Using all available VLA observations at X-band (8.0–12.0\,GHz), we produce the deepest radio map toward this system to date, with a noise level of 780\,nJy\,beam$^{-1}$. The source \mwc is detected with $S_\nu=4.6\pm0.8\,\mu$Jy and a spectral index of $\alpha=1.2\pm1.8$, derived by sub-band imaging. The radio and X-ray luminosity ratio of \mwc is consistent with both the fundamental plane of accreting black holes and with the G\"udel-Benz relation for magnetically active stars, leaving the emission mechanism ambiguous. 
The optical astrometric results of \mwc indicate a peculiar velocity of $11.2\pm2.3$\,\kms, discarding it as a runaway star. Its current location, 442\,pc below the Galactic plane, implies a vertical travel time incompatible with the lifetime of a B1.5-type star. Moreover, the agreement between observed and expected motion in all three velocity components argues against a deceleration scenario, suggesting that \mwc likely formed in situ at high Galactic latitude.
We carried out maximum likelihood analysis of \fermilat data, but cannot report a significant detection of \gammaray emission from this source. These results reinforce recent evidence that challenge the black hole companion interpretation, and favor a non-BH compact object such as a white dwarf or neutron star.
}


\maketitle
%

\section{Introduction}
Binary systems composed of a massive star and a compact object, i.e., a black hole (BH) or a neutron star (NS), are known as high-mass X-ray binaries (HMXBs). {These systems are important laboratories for studying accretion physics, compact object formation, and high-energy processes in stellar environments \citep[e.g.,][]{mirabel2007}. In classical HMXBs, accretion from the stellar wind or decretion disk of the massive star onto the compact object powers the observed X-ray and sometimes radio emission. If the compact object is a BH or an accreting NS, the high-energy emission is typically driven by accretion processes. In contrast, gamma-ray binaries, an energetic subclass of HMXBs, are generally powered by the interaction between the relativistic wind of a young pulsar and the circumstellar material of a Be star \citep{Dubus2013}. These systems exhibit orbital modulation across the electromagnetic spectrum, with gamma-ray emission not driven by accretion, but rather by shock acceleration in the pulsar-disk interaction zone. Understanding the energy source in a given system is thus crucial for constraining the nature of the compact companion.}
However, only a handful of HMXBs have been firmly identified as \gammaray binaries, with a few others considered as candidates. One such candidate, and the subject of this article, is \mwc (also known as HD~215227) {a binary system in a circular orbit with a period of ${59.028 \pm 0.002}$\,days \citep{Janssens2023}.  The primary is a Be star of spectral type B1.5--B2 III, and its projected rotational velocity (${v \sin i}$) has been measured to be approximately 330\,km\,s$^{-1}$ \citep{Casares2014}. The nature of the companion, as discussed below, has been a subject of debate during the last decade. The inclination of the system is estimated to lie between 30$^\circ$ and 80$^\circ$ based on modeling of the optical light curve and emission line profiles.}

\citet{Lucarelli2010} reported the detection of a new and previously unidentified point-like \gammaray source from maximum likelihood analysis of AGILE data ``at a significance level above $5\sigma$'' at the Galactic position $(l,b)=(100\rlap{.}^\circ2,-12\rlap{.}^\circ2)$ with a circular error of radius $0\rlap{.}^\circ6$. The AGILE source properties were consistent with it being a high-mass binary, pulsar, quasar or supernova, but none of these objects were found in the area of the source. \citet{Williams2010} noted that the Be star \mwc falls in the position of the AGILE source and concluded that it is likely a binary star with a compact object as companion and, consequently, a strong candidate counterpart of the \gammaray source. These authors also noted that because of the large Galactic latitude position of \mwc, it was possible that the supernova event that formed the compact companion could have kicked out the binary from the Galactic plane, making it a runaway star. However, the peculiar motions they found were inconclusive, given their large errors. A later study by \citet{Casares2014} estimated a mass range from 3.8 to 6.9\,M$_\odot$ for the compact companion in \mwc, claiming it to be the first case of a Be star in a binary with a black hole. Emerging evidence, however, is starting to disagree with these { results. In particular,} \citet{Janssens2023} used a similar approach as that employed by \citet{Casares2014} and found that the companion is in the range from 0.6 to 2.4\,M$_\odot$, discarding the BH companion scenario and favoring instead either a white dwarf, a neutron star, or a sub-dwarf O star (sdO) as the companion. {Preliminary findings from \citet{Rivinius2024}, presented at a recent conference proceedings, also support this trend, although a detailed peer-reviewed analysis is still pending.}

The association of \gammaray emission with the source \mwc has been challenged by the fact that the initial detection with the AGILE satellite, reported in {The Astronomer's Telegram} by \citet{Lucarelli2010}, could not be confirmed by independent observations (e.g., with the \fermilat). \citet{Alexander2015} revisited the AGILE observation by carrying out a maximum likelihood analysis. They confirmed the detection of the \gammaray source by \citet{Lucarelli2010}, albeit at a lower significance level of $3\sigma$. These authors also analyzed photon data from the \fermilat, but did not report any detection of \gammaray flux from \mwc{} in these data. Instead, they report the discovery of a previously undetected \gammaray source, which has entered the 3FGL \fermilat{} source catalog as 3FGL~J2201.7+5047 \citep{Acero2015}, but is not included in the latest version of the 4FGL catalog anymore \citep{Ballet2023}. A confirmation of the $5\sigma$ AGILE detection of \mwc was provided by \citet{Munar-Adrover2016}, along with several other events at a lower significance level ($>3\sigma$). There is, however, not any relation of these events to the orbital phase of \mwc. Also an analysis of \fermilat data carried out by these authors did not result in any significant detection, analyzing the full energy range of the \fermilat, i.e., 0.1-300\,GeV.

In this work, we revisit the nature of \mwc by conducting a multi-wavelength analysis using archival observations from the VLA, \textit{Gaia}, and \fermilat. Our aim is to reassess previous claims regarding the system's radio properties, kinematics, and possible \gammaray emission, in light of improved data and updated analysis tools. Through this comprehensive approach, we aim to clarify the physical nature of the compact companion and the high-energy characteristics of the system.


\section{Data and Analysis Methods}\label{sec:data}

In this section, we describe the datasets and analysis procedures used to investigate the properties of \mwc. Our study is based on archival observations from the {\it Karl G. Jansky} Very Large Array (VLA), the \textit{Gaia} satellite, and \fermilat. The details of the data processing and analysis techniques are described in the following subsections.

\subsection{VLA data analysis}

\begin{figure*}
  \includegraphics[width=0.99\linewidth]{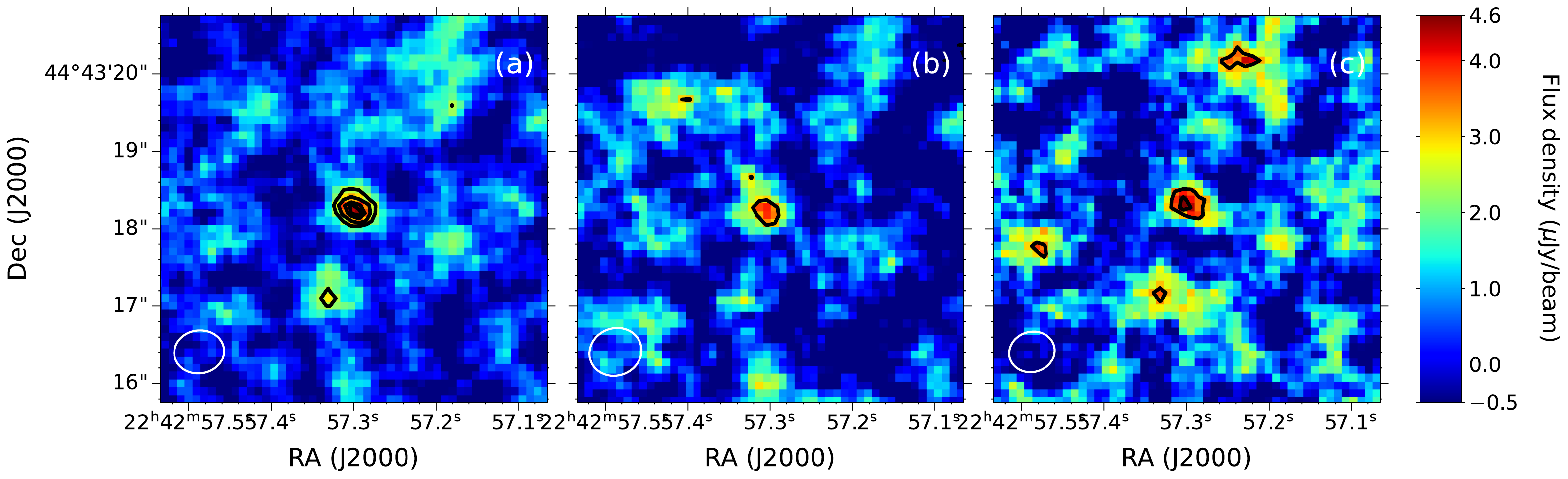}
  \caption{
    \mwc as detected by combining all eight observed epochs in the X-band with the VLA. 
    (a) Image of the full X-band.
    (b) Image of the LSB of the X-band, from 8.0 to 10.0 GHz.
    (c) Image of the USB of the X-band, from 10.0 to 12.0 GHz.
    Contour levels are --3.0, 3.0, 4.0, 5.0, and 5.4 the noise level of each image as listed in Table~\ref{tab:Nrad}. The synthesized beam size, also as listed in Table~\ref{tab:Nrad}, is shown as a white ellipse in the bottom-left of~each~panel.}
  \label{fig:full}
\end{figure*}

The first radio detection of \mwc was reported by \citet[][\citetalias{Dzib2015} hereafter]{Dzib2015} based on observations with the VLA in its B configuration. The instrument setup covered the full X-band (8 to 12~GHz) in semi-continuum mode.  Seven observations were carried out from February to April 2015, covering different orbital phases of the binary system. The duration of each observing session was two hours, which is very short compared to the $\sim 60$~days orbital period of the system.
The source was detected in their first epoch with a flux density of $14.2 \pm 2.9\,\mu$Jy, { while a}nalysis of the individual other six epochs yielded only upper limits (see their Table~1). By combining these six epochs (2-7), \citetalias{Dzib2015} determined a quiescent flux level of $3.7 \pm 1.4\,\mu$Jy. 

In July 2015, the source \mwc was observed again with the VLA, this time by \citet[][\citetalias{Ribo2017} hereafter]{Ribo2017}. These authors used the VLA in its A configuration and also observed the full X-band in semi-continuum mode. This observing session lasted for six hours. They reported an unresolved source with a flux density of $\unit[3.5 \pm 1.1]{\mu Jy}$, consistent with the result found by \citetalias{Dzib2015}. 

To improve the overall sensitivity level on VLA images, we here proceed to combine the observations by \citetalias{Dzib2015} and \citetalias{Ribo2017}. This approach is valid because of the consistency in instrumental setup and also because the source is unresolved, minimizing dependence on the observed angular resolution (i.e., the VLA configuration).

The data were calibrated using the CASA software \citep{CASA2022} following the calibration schemes described by \citetalias{Dzib2015} and \citetalias{Ribo2017}. The imaging process was also carried out with CASA, and produced by combining the eight calibrated data sets. It used a natural weighting scheme and with square pixels of 0\rlap{.}$''$1 per side. Three final images are produced: (i) the full observed band, (ii) a lower side band (LSB) in the range from 8.0 to 10 GHz, and  (iii) an upper side band (USB) covering the range from 10.0 to 12.0 GHz. 
During the cleaning process, the presence of a bright radio quasar at the west side of \mwc{} was taken into account \citep[see also][]{Moldon2012}. 
The resulting images are presented in Fig.~\ref{fig:full}.

\subsection{Optical astrometry}\label{sec3:gaia}

\mwc was identified in the \textit{Gaia} DR3 catalog under the ID code 1982359580155628160 \citep{gaia2021}. 
The measured parallax of $0.4860\pm0.0185$\,mas is equivalent to a distance of $2.06\pm0.08$\,kpc. This value represents a significant improvement over the commonly adopted distance for the system of $2.6\pm1.0$\,kpc \citep{Williams2010,Janssens2023}.
The \textit{Gaia} proper motions are $\mu_\alpha\cdot\cos{(\delta)}=-3.478\pm0.016$ and $\mu_\delta=-3.159\pm0.017$\,mas\,yr$^{-1}$.
In Galactic coordinates these proper motions are  $\mu_\ell\cdot\cos{(b)}=-4.698\pm0.018$ and $\mu_b=-0.851\pm0.017$\,mas\,yr$^{-1}$.
Additionally, the systemic radial velocity of \mwc is $-14.1\pm2.1$\,km\,s$^{-1}$ \citep{Casares2014}.

To investigate whether \mwc exhibits evidence of high peculiar motion, as previously suggested by \citet{Williams2010}, we analyze its space motion relative to expectations from Galactic rotation. Although \citet{fortin2022} previously examined the motion of \mwc from \textit{Gaia} results, their study focused on interactions with large-scale Galactic structures and did not discuss the runaway hypothesis.

The expected proper motion of \mwc can be estimated by adopting a simplified Galactic rotation model, in which stars in the Galactic disk move in circular orbits around the Galactic center. The coordinate system of this model is centered on the Galactic Center (GC), the $x$-axis runs from the Sun to the GC, the $y$-axis is in the direction of Galactic rotation, and the $z$-axis is towards the north Galactic pole. We assume a local standard of rest (LSR) speed of 254\,km\,s$^{-1}$ \citep{reid2009}, the Sun located at a distance of 8.4\,kpc from the Galactic center \citep{reid2009}. The solar motion relative to the LSR is taken to be $(U,\, V,\, W)=(11.10,12.24,7.25)$\,\kms \citep{schonrich2010}.

Under these assumptions, the coordinates and heliocentric distance of \mwc{}, we found a Galactic location of $(x,\,y,\,z)=(-8755.4\pm13.8,\,1980.3\pm76.9,\,-442.3\pm17.3)$\,pc. Based on this location, the expected tangential proper motions of a star in this position are $(\mu_\ell\cdot\cos{(b)},\,\mu_b)[{\rm expected}]=(-4.30\pm0.19,-1.12\pm0.15)$\,mas\,yr$^{-1}$, and with a systemic radial velocity of $-24.3\pm1.3$\,\kms. 
Comparing these values with the values from optical studies, we found that { the peculiar velocities of} \mwc  are $(v_{\ell},\,v_{b},v_{\rm rad})_{\rm pec.}=(-3.9\pm1.8$,\,$+2.5\pm1.3,\,+10.2\pm2.5)$\,\kms, { corresponding to a total peculiar velocity of $11.2\pm2.3$\,\kms}. 


{An additional analysis was carried out to check the velocity difference of \mwc with respect to stars in its surroundings. Then, we used the data archive of {\it Gaia} and searched for stars within $10'$ around the position of \mwc and with parallaxes between 
0.455\,mas (2.2\,kpc) and 0.526\,mas (1.9\,kpc). To ensure high-quality astrometric results, we also constrained the results
to stars with Renormalized Unit Weight Error parameter (RUWE\footnote{The RUWE parameter is a quality indicator of \textit{Gaia}
astrometric results. A low RUWE value (close to 1.0) suggests a reliable fit, while a large value ($>1.4$) may indicate source 
complexities, like multiplicity or other source of errors.}) $<1.4$ and parallax signal-to-noise ratio larger than 5. We found 
62 stars with these criteria. Our analysis shows that the mean proper motions are --3.13 and --3.63\,mas\,yr$^{-1}$ in R.A. and Decl., 
respectively. Then, the peculiar proper motions of MWC\,656 with respect to its surrounding stars are --0.35 and 0.04\,mas\,yr$^{-1}$ 
equivalent to velocities of --3.4 and 0.4\,km\,s$^{-1}$.
}

\subsection{\fermilat{} data analysis} \label{sec:fermi}

The Large Area Telescope (LAT) onboard the \textit{Fermi} satellite (\fermilat{}) has been taking data in all-sky monitoring mode since August 5, 2008 \citep{Atwood2009}. For the analysis presented here we used data until { September 18, 2025}, which is a total data set of { 17.1}~years. We used the {\it Fermitools}\footnote{Available from \url{https://github.com/fermi-lat/Fermitools-conda}} version~2.0.8 together with {\it Fermitools-data} version~0.18. We downloaded the \fermilat{} Pass-8 photon data from the \fermi{} science support center\footnote{\url{https://fermi.gsfc.nasa.gov/ssc/data/access/}}. 

The data included photons from a circle of radius 15\textdegree{} around the position of \mwc and span an energy range from 0.1 to 300~GeV. We use the script \texttt{make4FGLxml}\footnote{Downloaded from the user contributions section of the \fermi science support center: \url{https://fermi.gsfc.nasa.gov/ssc/data/analysis/user/}} to generate a model from the 4FGL \fermilat source catalog \citep{Abdollahi2020}. This initial model includes all sources that yield a $5\sigma$ significance over the integration time of the catalog. {The fact that \mwc is not included in the catalog shows that this object is not detected as a source of \gammaray emission when combining all \fermi data. This excludes \mwc as a persistent \gammaray source, but transient emission is still a possibility, which we investigated with the approach described in the following.}

The results presented in \citet{Munar-Adrover2016} suggest that the putative \gammaray source associated with \mwc has a power-law spectral shape with a photon index $\gamma = 2.3 \pm 0.2$ and is only detected until an energy of 1~GeV (see their Fig.~3). Based on these findings, we added to the model a point source at the position of \mwc with a power law spectral shape,
\begin{equation}
        \frac{\mathrm{d}N}{\mathrm{d}E} = N_0\left(\frac{E}{E_0}\right)^{-\gamma},
\end{equation}
with prefactor $N_0$, photon index $\gamma$, and energy scale $E_0$. We fixed the photon index to the value $\gamma = 2.3$ and left only the prefactor $N_0$ free for the fit. Furthermore, the transient detections listed in Table~1 of \citet{Munar-Adrover2016} suggest that these events of \gammaray flares have a typical duration of 1-3~days. With this knowledge in mind, we divided the { 17.1}~years of available photon data into time bins of 2.9514~days, which is exactly $P_{\rm orb}$/20, adopting the updated orbital period of 59.028~d by \citet{Janssens2023}. The aim of this choice is to simplify any analysis of possible orbital phase dependency of the results. We then performed an unbinned likelihood analysis in each of these 1931 time bins separately. We restricted the photon energy range to $E = 0.1 - 1.0$\,GeV. Normalization factors of all sources within a radius of 5\textdegree{} were left free for the fit and their spectral parameters were fixed to the catalog values. We used \texttt{gll\_iem\_v07.fits} as the model for the Galactic diffuse emission and \texttt{iso\_P8R3\_SOURCE\_V3\_v1.txt} as the template for the isotropic background emission. Prefactor and normalization were left free for the fit for these two components, respectively. We do not detect the source \mwc{} in any of these time bins. The largest test statistic (TS) value that we obtain is 0.31, which is not even close to a value of 9, which would approximately correspond to a $3\sigma$ detection \citep{Abdo2009}. { We also divided the Fermi data into ten orbital phase bins, according to the orbital period determined by \citet{Janssens2023} (i.e., the phase-bins had a width of 5.9028~d each) and performed the likelihood analysis in each of these bins. The source was not detected in any of these bins either.}


\section{Results and Discussions} \label{sec:results}

In this section, we present and discuss the results of our multi-wavelength study of \mwc{}, combining radio continuum imaging with optical astrometry. 

\subsection{Radio properties of \mwc{}}

The combined dataset used to produce a new image of \mwc{} resulted in a detection with a { flux density of ${4.6\pm0.8\,\mu}$Jy and} signal-to-noise ratio (S/N) of 5.9, representing the most significant radio detection of the source to date.  { To test the robustness of the detection, we re-imaged the data excluding the first epoch from 2015, which corresponds to the brightest single detection reported by \citetalias{Dzib2015}. The source is still clearly detected, with a flux density of ${4.3 \pm 0.8\,\mu}$Jy and a noise level of ${0.84\,\mu}$Jy\,beam$^{-1}$. These values are consistent with the full dataset, confirming that the detection is not driven by a single bright epoch.} \mwc\ is also detected in the images of the two defined sub-bands, each with S/N$\sim4.0$. Details of the radio detections are given in Table~\ref{tab:Nrad}.

We assume that the source flux density at these frequencies can be modeled as a power-law function of the form $S_\nu\propto\nu^{+\alpha}$, where $\alpha$ is the spectral index. Using the information provided by the two sub-bands, with central frequencies of 9.0\,GHz and 11.0\,GHz, we estimated the spectral index to be $\alpha=1.2\pm1.8$.

{We emphasize that the uncertainty of the spectral index is large ($\pm 1.8$), such that optically thin synchrotron emission (e.g., $\alpha \approx -0.6$) lies well within $1\sigma$. Therefore, no specific emission mechanism can be confidently favored or excluded based on the spectral slope alone. The nominal value, however, is marginally consistent with a flat or mildly inverted spectrum.} In a microquasar scenario, such a radio spectrum is typical of accreting BHs in a low/hard X-ray state \citep{Fender2004}. The X-ray properties of \mwc{} found by \citet{Ribo2017} are indeed compatible with a BH in deep quiescence, as these authors point out. { However, this interpretation is increasingly disfavored in light of recent \mwc mass estimates.}

If the compact star is not a BH, as recent observations suggest, then there is still the possibility of an accreting neutron star (NS), which could explain the observed radio and X-ray properties. However, as \citet{Janssens2023} point out, the radio emission from Be + NS binaries has not yet been explored in the regime of such low X-ray luminosities \citep{Eijnden2021}. { If the radio emission from \mwc is confirmed to originate in a jet, then, according to \citet{Massi2008}, this would put constraints on the magnetic field of a neutron star in this system (see, however, \citealt{Eijnden2021}).}


The monochromatic radio luminosity of \mwc at the observed frequency is $L_{\nu}=2.6\times10^{16}$\,erg\,s$^{-1}$\,Hz$^{-1}$, adopting a distance of 2.06\,kpc. The total radio luminosity at the observed bandwidth can be estimated by assuming a flat spectrum to be  $L_{\rm radio}= 1.04\times10^{26}$\,erg\,s$^{-1}$. On the other hand, the X-ray luminosity reported by \citetalias{Ribo2017} translates to $L_{X}=1.95\times10^{30}$\,erg\,s$^{-1}$, after rescaling it to the \mwc distance of 2.06\,kpc assumed in this work. These values remain consistent with \mwc lying in the so-called fundamental plane of accreting BHs \citep{Merloni2003,gallo2006,plotkin2017}, but in the faint edge, as previously noted by \citetalias{Dzib2015} and \citetalias{Ribo2017}. 

However, we note that the radio--X-ray luminosity ratio of ${(L_{X} / L_{\nu})}_{\rm MWC656} = 7.5 \times 10^{13}$\,Hz$^{-1}$ is also broadly consistent with values reported for magnetically active young stars, as described by the Güdel-Benz relation \citep{gb1993, dzib2015b, yanza2022}. Given the current data, the origin of the radio emission remains ambiguous, and both accretion- and magnetically-driven scenarios should be considered.

\begin{table}[]
    \begin{center}
    \caption{\mwc{} and VLA image parameters.}\label{tab:Nrad}
    \begin{tabular}{cccc}
    \hline\hline
      $\nu$-range   & Synthesized beam         & Noise                 & $S_\nu$   \\
        (GHz)           & ($''\,\times\,''$; $^\circ$) & ($\mu$Jy bm$^{-1}$) & ($\mu$Jy) \\
        \hline
        8.0 -- 12.0     & $0.64\times0.55; -82$ & 0.78  & $4.6\pm0.8$ \\
        8.0 -- 10.0     & $0.68\times0.61; -70$ & 1.10  & $4.0\pm1.1$ \\
        10.0 -- 12.0    & $0.59\times0.52; -74$ & 1.20  & $5.1\pm1.2$ \\
        \hline
    \end{tabular}
    \end{center}
    \noindent {Notes: $\Delta\nu$ is the frequency range used to produce the image. The synthesized beam corresponds to the angular resolution. $S_\nu$ is the \mwc{} measured flux density.}
\end{table}

\subsection{Astrometry}

{ Runaway stars are defined as stars moving at high velocities, usually above 30 to 40\,\kms, relative to the local standard of rest velocity \citep{moffat1998, hoogerwerf2001}. In the case of \mwc, we derived a total peculiar velocity of $11.2\pm2.3$\,\kms, well below this threshold.  Furthermore, the peculiar velocity of \mwc relative to the local stars is 3.4\,\kms. Both results confirm that the system does not meet the conventional criteria for classification as a runaway star.}

The celestial coordinates and heliocentric distance of \mwc indicate that it is at 442\,pc below the Galactic plane.  Assuming that the system originated in the Galactic plane and has been continuously moving away at its present vertical velocity, derived from the Galactic proper motion $\mu_b = -0.851 \pm 0.017$\,mas\,yr$^{-1}$ (corresponding to $v_b = 8.30 \pm 0.17$\,km\,s$^{-1}$), it would require approximately 52 million years to reach its current altitude.  This timescale is 3.5 times the main-sequence lifetime of a B1.5-type star, which is typically less than 15 million years \citep[e.g., ][]{ekstrom2012,brott2011}. This discrepancy suggests that a birth in the Galactic plane is unlikely. Moreover, the fact that the system's velocity components in Galactic longitude, latitude, and radial directions all agree with expectations from Galactic rotation makes a deceleration scenario unlikely, as it would require simultaneous energy loss along all axes.

\citet{fortin2022} found no evidence of interaction between the \mwc{} system and any spiral arms or open clusters, and suggested that it may have formed in isolation. Our results support this interpretation, though with a different emphasis. The proper motion and radial velocity of \mwc{} are in good agreement with the predictions of a standard Galactic rotation model. This implies that the system has likely not migrated far from its birthplace, and raises the possibility that it formed in situ at a significant height above the Galactic plane (i.e., outside the thin disk).
 Compared to earlier estimates with large uncertainties \citep[e.g.,][]{Williams2010}, our use of Gaia DR3 proper motions and updated radial velocity enables a significantly more precise derivation of the system's peculiar motion. This precision effectively rules out a runaway scenario and reinforces the conclusion that the system likely formed at its current position, a result not accessible in earlier studies.

\subsection{The nature of the \mwc{} companion}

The potential in-situ formation of \mwc{} at a high Galactic latitude (442\,pc below the plane) invites further reflection on the evolutionary history of the system. Star formation far from the Galactic plane is relatively rare and typically limited to low-density regions, with few massive stellar clusters. These conditions are unfavorable for the formation of very massive stars, and thus for BH progenitors. In contrast, the long-term evolution of intermediate-mass binaries—including episodes of mass transfer and common envelope evolution—may more plausibly result in the formation of white dwarfs or neutron stars. Moreover, the low peculiar velocity of \mwc{} argues against a strong natal kick, which would be more characteristic of a supernova that forms a NS or BH. Taken together, these environmental and kinematic arguments suggest that a compact companion may indeed be present, but likely not a BH.

Additional clues arise from the observed radio and X-ray emission. While such emission is often interpreted as a signature of accretion onto a compact object, the radio--X-ray luminosity ratio of \mwc{} is also consistent with the G\"udel–Benz relation, which characterizes magnetically active stars \citep{gb1993, dzib2015b, yanza2022}. This opens the possibility that the emission originates from a low-mass pre-main-sequence or main-sequence star with a magnetically heated corona, rather than from an accreting compact object. Thus, the presence of X-ray and radio emission does not conclusively establish the nature of the companion.

However, not all companion types are equally consistent with the available constraints. The orbital solution implies a companion mass between 0.6 and 2.4\,M$_\odot$ \citep{Janssens2023}, and no optical contribution from a secondary is detected \citep{Casares2014, Janssens2023}. These conditions disfavor mid- or high-mass main-sequence stars. A low-mass stellar companion remains viable but would require a specific inclination and evolutionary scenario to remain undetected. A white dwarf or neutron star, both optically faint and radio/X-ray-active under certain conditions, offers a more natural explanation. While the identity of the companion cannot yet be definitively determined, the combined astrometric, photometric, and radiative evidence favors a non-black-hole compact object.

\section{Conclusions} \label{sec:conclusions}

In this work, we have revisited the nature of the \mwc{} system with radio interferometry, optical astrometry from \textit{Gaia}, and high-energy \gammaray{} data from \fermilat{}.  These are our conclusions:
\begin{enumerate}
    \item{
        Combining the VLA data sets from \citet{Dzib2015} and \citet{Ribo2017} improves the significance of the radio detection to $5.9\sigma$, which is thereby confirmed and strengthened.
    }
    \item The estimated spectral index ($S \propto \nu^{+\alpha}$) of the radio emission is $\alpha=1.2\pm1.8$. Although uncertain, {its nominal value tentatively} indicates a flat or inverted spectrum, which would be compatible with an accreting compact object (NS or BH) in quiescence. However,  given the large uncertainty of $\alpha$, the origin of the radio emission remains ambiguous.
    \item The comparisons of radio and X-ray luminosities are consistent with both the fundamental plane of accreting BHs and with the G\"udel-Benz relation for magnetically active stars. Then, the relation of these wavelengths is also ambiguous in this case. 
    \item The measured proper motions and radial velocity of \mwc{} are consistent with Galactic rotation and imply a peculiar velocity well below the threshold for runaway stars. This argues against a runaway origin for the system.
    \item{A maximum likelihood analysis of \fermilat{} data does not result in any detection of \gammaray{} emission from the region where \mwc{} is located.    }
    
\end{enumerate}

Although \mwc{} was previously proposed as the first Be+BH binary, recent evidence has cast doubt on this scenario. Our results further support this re-evaluation, particularly by excluding a runaway origin based on the kinematics of the system and by finding no evidence for associated $\gamma$-ray emission in the \fermilat{} data.

Nevertheless, the large vertical distance of \mwc{} (442\,pc) from the Galactic plane remains puzzling, especially given the lack of nearby regions of star formation at such Galactic latitudes. The origin of \mwc{} thus remains unclear and may point toward atypical or isolated formation pathways outside the thin disk.

\begin{acknowledgements}
We thank the anonymous referee for their thoughtful comments that helped improve this paper.
We thank M. Massi and E. Ros for useful discussions and suggestions. 
S.A.D. acknowledges the M2FINDERS project from the European Research
Council (ERC) under the European Union's Horizon 2020 research and innovation programme
(grant No 101018682).
This work has made use of public Fermi data obtained from the High Energy Astrophysics Science Archive Research Center (HEASARC), provided by NASA Goddard Space Flight Center. The computational results presented have been achieved [in part] using the Vienna Scientific Cluster (VSC). This research was funded by the Austrian Science Fund (FWF) [P31625].
The National Radio Astronomy Observatory is a facility of the National Science Foundation operated under cooperative agreement by Associated Universities, Inc.
\end{acknowledgements}

\bibliographystyle{aa_url} 
\bibliography{mwc656.bib} 

\begin{thebibliography}{35}
\expandafter\ifx\csname natexlab\endcsname\relax\def\natexlab#1{#1}\fi

\bibitem[{{Abdo} {et~al.}(2009){Abdo}, {Ackermann}, {Ajello}, {Atwood}, {Axelsson}, {Baldini}, {Ballet}, {Band}, {Barbiellini}, {Bastieri}, {Battelino}, {Baughman}, {Bechtol}, {Bellazzini}, {Berenji}, {Bignami}, {Blandford}, {Bloom}, {Bonamente}, {Borgland}, {Bouvier}, {Bregeon}, {Brez}, {Brigida}, {Bruel}, {Burnett}, {Caliandro}, {Cameron}, {Caraveo}, {Casandjian}, {Cavazzuti}, {Cecchi}, {Charles}, {Chekhtman}, {Cheung}, {Chiang}, {Ciprini}, {Claus}, {Cohen-Tanugi}, {Cominsky}, {Conrad}, {Corbet}, {Costamante}, {Cutini}, {Davis}, {Dermer}, {de Angelis}, {de Luca}, {de Palma}, {Digel}, {Dormody}, {do Couto e Silva}, {Drell}, {Dubois}, {Dumora}, {Farnier}, {Favuzzi}, {Fegan}, {Ferrara}, {Focke}, {Frailis}, {Fukazawa}, {Funk}, {Fusco}, {Gargano}, {Gasparrini}, {Gehrels}, {Germani}, {Giebels}, {Giglietto}, {Giommi}, {Giordano}, {Glanzman}, {Godfrey}, {Grenier}, {Grondin}, {Grove}, {Guillemot}, {Guiriec}, {Hanabata}, {Harding}, {Hartman}, {Hayashida}, {Hays}, {Healey}, {Horan}, {Hughes}, {J{\'o}hannesson},
  {Johnson}, {Johnson}, {Johnson}, {Johnson}, {Kamae}, {Katagiri}, {Kataoka}, {Kawai}, {Kerr}, {Kn{\"o}dlseder}, {Kocevski}, {Kocian}, {Komin}, {Kuehn}, {Kuss}, {Lande}, {Latronico}, {Lee}, {Lemoine-Goumard}, {Longo}, {Loparco}, {Lott}, {Lovellette}, {Lubrano}, {Madejski}, {Makeev}, {Marelli}, {Mazziotta}, {McConville}, {McEnery}, {McGlynn}, {Meurer}, {Michelson}, {Mitthumsiri}, {Mizuno}, {Moiseev}, {Monte}, {Monzani}, {Moretti}, {Morselli}, {Moskalenko}, {Murgia}, {Nakamori}, {Nolan}, {Norris}, {Nuss}, {Ohno}, {Ohsugi}, {Omodei}, {Orlando}, {Ormes}, {Ozaki}, {Paneque}, {Panetta}, {Parent}, {Pelassa}, {Pepe}, {Pesce-Rollins}, {Piron}, {Porter}, {Poupard}, {Rain{\`o}}, {Rando}, {Ray}, {Razzano}, {Rea}, {Reimer}, {Reimer}, {Reposeur}, {Ritz}, {Rochester}, {Rodriguez}, {Romani}, {Roth}, {Ryde}, {Sadrozinski}, {Sanchez}, {Sander}, {Saz Parkinson}, {Scargle}, {Schalk}, {Sellerholm}, {Sgr{\`o}}, {Shaw}, {Shrader}, {Sierpowska-Bartosik}, {Siskind}, {Smith}, {Smith}, {Spandre}, {Spinelli}, {Starck}, {Stephens},
  {Strickman}, {Strong}, {Suson}, {Tajima}, {Takahashi}, {Takahashi}, {Tanaka}, {Thayer}, {Thayer}, {Thompson}, {Tibaldo}, {Tibolla}, {Torres}, {Tosti}, {Tramacere}, {Uchiyama}, {Usher}, {Van Etten}, \& {Vilchez}}]{Abdo2009}
{Abdo}, A.~A., {Ackermann}, M., {Ajello}, M., {et~al.} 2009, \href{http://dx.doi.org/10.1088/0067-0049/183/1/46}{\color{magenta}\apjs}, \href{https://ui.adsabs.harvard.edu/abs/2009ApJS..183...46A}{183, 46}

\bibitem[{{Abdollahi} {et~al.}(2020){Abdollahi}, {Acero}, {Ackermann}, {Ajello}, {Atwood}, {Axelsson}, {Baldini}, {Ballet}, {Barbiellini}, {Bastieri}, {Becerra Gonzalez}, {Bellazzini}, {Berretta}, {Bissaldi}, {Blandford}, {Bloom}, {Bonino}, {Bottacini}, {Brandt}, {Bregeon}, {Bruel}, {Buehler}, {Burnett}, {Buson}, {Cameron}, {Caputo}, {Caraveo}, {Casandjian}, {Castro}, {Cavazzuti}, {Charles}, {Chaty}, {Chen}, {Cheung}, {Chiaro}, {Ciprini}, {Cohen-Tanugi}, {Cominsky}, {Coronado-Bl{\'a}zquez}, {Costantin}, {Cuoco}, {Cutini}, {D'Ammando}, {DeKlotz}, {de la Torre Luque}, {de Palma}, {Desai}, {Digel}, {Di Lalla}, {Di Mauro}, {Di Venere}, {Dom{\'\i}nguez}, {Dumora}, {Fana Dirirsa}, {Fegan}, {Ferrara}, {Franckowiak}, {Fukazawa}, {Funk}, {Fusco}, {Gargano}, {Gasparrini}, {Giglietto}, {Giommi}, {Giordano}, {Giroletti}, {Glanzman}, {Green}, {Grenier}, {Griffin}, {Grondin}, {Grove}, {Guiriec}, {Harding}, {Hayashi}, {Hays}, {Hewitt}, {Horan}, {J{\'o}hannesson}, {Johnson}, {Kamae}, {Kerr}, {Kocevski}, {Kovac'evic'},
  {Kuss}, {Landriu}, {Larsson}, {Latronico}, {Lemoine-Goumard}, {Li}, {Liodakis}, {Longo}, {Loparco}, {Lott}, {Lovellette}, {Lubrano}, {Madejski}, {Maldera}, {Malyshev}, {Manfreda}, {Marchesini}, {Marcotulli}, {Mart{\'\i}-Devesa}, {Martin}, {Massaro}, {Mazziotta}, {McEnery}, {Mereu}, {Meyer}, {Michelson}, {Mirabal}, {Mizuno}, {Monzani}, {Morselli}, {Moskalenko}, {Negro}, {Nuss}, {Ojha}, {Omodei}, {Orienti}, {Orlando}, {Ormes}, {Palatiello}, {Paliya}, {Paneque}, {Pei}, {Pe{\~n}a-Herazo}, {Perkins}, {Persic}, {Pesce-Rollins}, {Petrosian}, {Petrov}, {Piron}, {Poon}, {Porter}, {Principe}, {Rain{\`o}}, {Rando}, {Razzano}, {Razzaque}, {Reimer}, {Reimer}, {Remy}, {Reposeur}, {Romani}, {Saz Parkinson}, {Schinzel}, {Serini}, {Sgr{\`o}}, {Siskind}, {Smith}, {Spandre}, {Spinelli}, {Strong}, {Suson}, {Tajima}, {Takahashi}, {Tak}, {Thayer}, {Thompson}, {Tibaldo}, {Torres}, {Torresi}, {Valverde}, {Van Klaveren}, {van Zyl}, {Wood}, {Yassine}, \& {Zaharijas}}]{Abdollahi2020}
{Abdollahi}, S., {Acero}, F., {Ackermann}, M., {et~al.} 2020, \href{http://dx.doi.org/10.3847/1538-4365/ab6bcb}{\color{magenta}\apjs}, \href{https://ui.adsabs.harvard.edu/abs/2020ApJS..247...33A}{247, 33}

\bibitem[{{Acero} {et~al.}(2015){Acero}, {Ackermann}, {Ajello}, {Albert}, {Atwood}, {Axelsson}, {Baldini}, {Ballet}, {Barbiellini}, {Bastieri}, {Belfiore}, {Bellazzini}, {Bissaldi}, {Blandford}, {Bloom}, {Bogart}, {Bonino}, {Bottacini}, {Bregeon}, {Britto}, {Bruel}, {Buehler}, {Burnett}, {Buson}, {Caliandro}, {Cameron}, {Caputo}, {Caragiulo}, {Caraveo}, {Casandjian}, {Cavazzuti}, {Charles}, {Chaves}, {Chekhtman}, {Cheung}, {Chiang}, {Chiaro}, {Ciprini}, {Claus}, {Cohen-Tanugi}, {Cominsky}, {Conrad}, {Cutini}, {D'Ammando}, {de Angelis}, {DeKlotz}, {de Palma}, {Desiante}, {Digel}, {Di Venere}, {Drell}, {Dubois}, {Dumora}, {Favuzzi}, {Fegan}, {Ferrara}, {Finke}, {Franckowiak}, {Fukazawa}, {Funk}, {Fusco}, {Gargano}, {Gasparrini}, {Giebels}, {Giglietto}, {Giommi}, {Giordano}, {Giroletti}, {Glanzman}, {Godfrey}, {Grenier}, {Grondin}, {Grove}, {Guillemot}, {Guiriec}, {Hadasch}, {Harding}, {Hays}, {Hewitt}, {Hill}, {Horan}, {Iafrate}, {Jogler}, {J{\'o}hannesson}, {Johnson}, {Johnson}, {Johnson}, {Johnson}, {Kamae},
  {Kataoka}, {Katsuta}, {Kuss}, {La Mura}, {Landriu}, {Larsson}, {Latronico}, {Lemoine-Goumard}, {Li}, {Li}, {Longo}, {Loparco}, {Lott}, {Lovellette}, {Lubrano}, {Madejski}, {Massaro}, {Mayer}, {Mazziotta}, {McEnery}, {Michelson}, {Mirabal}, {Mizuno}, {Moiseev}, {Mongelli}, {Monzani}, {Morselli}, {Moskalenko}, {Murgia}, {Nuss}, {Ohno}, {Ohsugi}, {Omodei}, {Orienti}, {Orlando}, {Ormes}, {Paneque}, {Panetta}, {Perkins}, {Pesce-Rollins}, {Piron}, {Pivato}, {Porter}, {Racusin}, {Rando}, {Razzano}, {Razzaque}, {Reimer}, {Reimer}, {Reposeur}, {Rochester}, {Romani}, {Salvetti}, {S{\'a}nchez-Conde}, {Saz Parkinson}, {Schulz}, {Siskind}, {Smith}, {Spada}, {Spandre}, {Spinelli}, {Stephens}, {Strong}, {Suson}, {Takahashi}, {Takahashi}, {Tanaka}, {Thayer}, {Thayer}, {Thompson}, {Tibaldo}, {Tibolla}, {Torres}, {Torresi}, {Tosti}, {Troja}, {Van Klaveren}, {Vianello}, {Winer}, {Wood}, {Wood}, {Zimmer}, \& {Fermi-LAT Collaboration}}]{Acero2015}
{Acero}, F., {Ackermann}, M., {Ajello}, M., {et~al.} 2015, \href{http://dx.doi.org/10.1088/0067-0049/218/2/23}{\color{magenta}\apjs}, \href{https://ui.adsabs.harvard.edu/abs/2015ApJS..218...23A}{218, 23}

\bibitem[{{Alexander} \& {McSwain}(2015)}]{Alexander2015}
{Alexander}, M.~J. \& {McSwain}, M.~V. 2015, \href{http://dx.doi.org/10.1093/mnras/stv400}{\color{magenta}\mnras}, \href{https://ui.adsabs.harvard.edu/abs/2015MNRAS.449.1686A}{449, 1686}

\bibitem[{{Atwood} {et~al.}(2009){Atwood}, {Abdo}, {Ackermann}, {Althouse}, {Anderson}, {Axelsson}, {Baldini}, {Ballet}, {Band}, {Barbiellini}, {Bartelt}, {Bastieri}, {Baughman}, {Bechtol}, {B{\'e}d{\'e}r{\`e}de}, {Bellardi}, {Bellazzini}, {Berenji}, {Bignami}, {Bisello}, {Bissaldi}, {Blandford}, {Bloom}, {Bogart}, {Bonamente}, {Bonnell}, {Borgland}, {Bouvier}, {Bregeon}, {Brez}, {Brigida}, {Bruel}, {Burnett}, {Busetto}, {Caliandro}, {Cameron}, {Caraveo}, {Carius}, {Carlson}, {Casandjian}, {Cavazzuti}, {Ceccanti}, {Cecchi}, {Charles}, {Chekhtman}, {Cheung}, {Chiang}, {Chipaux}, {Cillis}, {Ciprini}, {Claus}, {Cohen-Tanugi}, {Condamoor}, {Conrad}, {Corbet}, {Corucci}, {Costamante}, {Cutini}, {Davis}, {Decotigny}, {DeKlotz}, {Dermer}, {de Angelis}, {Digel}, {do Couto e Silva}, {Drell}, {Dubois}, {Dumora}, {Edmonds}, {Fabiani}, {Farnier}, {Favuzzi}, {Flath}, {Fleury}, {Focke}, {Funk}, {Fusco}, {Gargano}, {Gasparrini}, {Gehrels}, {Gentit}, {Germani}, {Giebels}, {Giglietto}, {Giommi}, {Giordano}, {Glanzman},
  {Godfrey}, {Grenier}, {Grondin}, {Grove}, {Guillemot}, {Guiriec}, {Haller}, {Harding}, {Hart}, {Hays}, {Healey}, {Hirayama}, {Hjalmarsdotter}, {Horn}, {Hughes}, {J{\'o}hannesson}, {Johansson}, {Johnson}, {Johnson}, {Johnson}, {Johnson}, {Kamae}, {Katagiri}, {Kataoka}, {Kavelaars}, {Kawai}, {Kelly}, {Kerr}, {Klamra}, {Kn{\"o}dlseder}, {Kocian}, {Komin}, {Kuehn}, {Kuss}, {Landriu}, {Latronico}, {Lee}, {Lee}, {Lemoine-Goumard}, {Lionetto}, {Longo}, {Loparco}, {Lott}, {Lovellette}, {Lubrano}, {Madejski}, {Makeev}, {Marangelli}, {Massai}, {Mazziotta}, {McEnery}, {Menon}, {Meurer}, {Michelson}, {Minuti}, {Mirizzi}, {Mitthumsiri}, {Mizuno}, {Moiseev}, {Monte}, {Monzani}, {Moretti}, {Morselli}, {Moskalenko}, {Murgia}, {Nakamori}, {Nishino}, {Nolan}, {Norris}, {Nuss}, {Ohno}, {Ohsugi}, {Omodei}, {Orlando}, {Ormes}, {Paccagnella}, {Paneque}, {Panetta}, {Parent}, {Pearce}, {Pepe}, {Perazzo}, {Pesce-Rollins}, {Picozza}, {Pieri}, {Pinchera}, {Piron}, {Porter}, {Poupard}, {Rain{\`o}}, {Rando}, {Rapposelli}, {Razzano},
  {Reimer}, {Reimer}, {Reposeur}, {Reyes}, {Ritz}, {Rochester}, {Rodriguez}, {Romani}, {Roth}, {Russell}, {Ryde}, {Sabatini}, {Sadrozinski}, {Sanchez}, {Sander}, {Sapozhnikov}, {Parkinson}, {Scargle}, {Schalk}, {Scolieri}, {Sgr{\`o}}, {Share}, {Shaw}, {Shimokawabe}, {Shrader}, {Sierpowska-Bartosik}, {Siskind}, {Smith}, {Smith}, {Spandre}, {Spinelli}, {Starck}, {Stephens}, {Strickman}, {Strong}, {Suson}, {Tajima}, {Takahashi}, {Takahashi}, {Tanaka}, {Tenze}, {Tether}, {Thayer}, {Thayer}, {Thompson}, {Tibaldo}, {Tibolla}, {Torres}, {Tosti}, {Tramacere}, {Turri}, {Usher}, {Vilchez}, {Vitale}, {Wang}, {Watters}, {Winer}, {Wood}, {Ylinen}, \& {Ziegler}}]{Atwood2009}
{Atwood}, W.~B., {Abdo}, A.~A., {Ackermann}, M., {et~al.} 2009, \href{http://dx.doi.org/10.1088/0004-637X/697/2/1071}{\color{magenta}\apj}, \href{https://ui.adsabs.harvard.edu/abs/2009ApJ...697.1071A}{697, 1071}

\bibitem[{{Ballet} {et~al.}(2023){Ballet}, {Bruel}, {Burnett}, {Lott}, \& {The Fermi-LAT collaboration}}]{Ballet2023}
{Ballet}, J., {Bruel}, P., {Burnett}, T.~H., {Lott}, B., \& {The Fermi-LAT collaboration}. 2023, \href{https://ui.adsabs.harvard.edu/abs/2023arXiv230712546B}{\href{http://dx.doi.org/10.48550/arXiv.2307.12546}{\color{magenta}arXiv e-prints}, arXiv:2307.12546}

\bibitem[{{Brott} {et~al.}(2011){Brott}, {de Mink}, {Cantiello}, {Langer}, {de Koter}, {Evans}, {Hunter}, {Trundle}, \& {Vink}}]{brott2011}
{Brott}, I., {de Mink}, S.~E., {Cantiello}, M., {et~al.} 2011, \href{http://dx.doi.org/10.1051/0004-6361/201016113}{\color{magenta}\aap}, \href{https://ui.adsabs.harvard.edu/abs/2011A&A...530A.115B}{530, A115}

\bibitem[{{CASA Team} {et~al.}(2022){CASA Team}, {Bean}, {Bhatnagar}, {Castro}, {Donovan Meyer}, {Emonts}, {Garcia}, {Garwood}, {Golap}, {Gonzalez Villalba}, {Harris}, {Hayashi}, {Hoskins}, {Hsieh}, {Jagannathan}, {Kawasaki}, {Keimpema}, {Kettenis}, {Lopez}, {Marvil}, {Masters}, {McNichols}, {Mehringer}, {Miel}, {Moellenbrock}, {Montesino}, {Nakazato}, {Ott}, {Petry}, {Pokorny}, {Raba}, {Rau}, {Schiebel}, {Schweighart}, {Sekhar}, {Shimada}, {Small}, {Steeb}, {Sugimoto}, {Suoranta}, {Tsutsumi}, {van Bemmel}, {Verkouter}, {Wells}, {Xiong}, {Szomoru}, {Griffith}, {Glendenning}, \& {Kern}}]{CASA2022}
{CASA Team}, {Bean}, B., {Bhatnagar}, S., {et~al.} 2022, \href{http://dx.doi.org/10.1088/1538-3873/ac9642}{\color{magenta}\pasp}, \href{https://ui.adsabs.harvard.edu/abs/2022PASP..134k4501C}{134, 114501}

\bibitem[{{Casares} {et~al.}(2014){Casares}, {Negueruela}, {Rib{\'o}}, {Ribas}, {Paredes}, {Herrero}, \& {Sim{\'o}n-D{\'\i}az}}]{Casares2014}
{Casares}, J., {Negueruela}, I., {Rib{\'o}}, M., {et~al.} 2014, \href{http://dx.doi.org/10.1038/nature12916}{\color{magenta}\nat}, \href{https://ui.adsabs.harvard.edu/abs/2014Natur.505..378C}{505, 378}

\bibitem[{{Dubus}(2013)}]{Dubus2013}
{Dubus}, G. 2013, \href{http://dx.doi.org/10.1007/s00159-013-0064-5}{\color{magenta}\aapr}, \href{https://ui.adsabs.harvard.edu/abs/2013A&ARv..21...64D}{21, 64}

\bibitem[{{Dzib} {et~al.}(2015{\natexlab{a}}){Dzib}, {Loinard}, {Rodr{\'\i}guez}, {Mioduszewski}, {Ortiz-Le{\'o}n}, {Kounkel}, {Pech}, {Rivera}, {Torres}, {Boden}, {Hartmann}, {Evans}, {Brice{\~n}o}, \& {Tobin}}]{dzib2015b}
{Dzib}, S.~A., {Loinard}, L., {Rodr{\'\i}guez}, L.~F., {et~al.} 2015{\natexlab{a}}, \href{http://dx.doi.org/10.1088/0004-637X/801/2/91}{\color{magenta}\apj}, \href{https://ui.adsabs.harvard.edu/abs/2015ApJ...801...91D}{801, 91}

\bibitem[{{Dzib} {et~al.}(2015{\natexlab{b}}){Dzib}, {Massi}, \& {Jaron}}]{Dzib2015}
{Dzib}, S.~A., {Massi}, M., \& {Jaron}, F. 2015{\natexlab{b}}, \href{http://dx.doi.org/10.1051/0004-6361/201526867}{\color{magenta}\aap}, \href{https://ui.adsabs.harvard.edu/abs/2015A&A...580L...6D}{580, L6}, [D15]

\bibitem[{{Ekstr{\"o}m} {et~al.}(2012){Ekstr{\"o}m}, {Georgy}, {Eggenberger}, {Meynet}, {Mowlavi}, {Wyttenbach}, {Granada}, {Decressin}, {Hirschi}, {Frischknecht}, {Charbonnel}, \& {Maeder}}]{ekstrom2012}
{Ekstr{\"o}m}, S., {Georgy}, C., {Eggenberger}, P., {et~al.} 2012, \href{http://dx.doi.org/10.1051/0004-6361/201117751}{\color{magenta}\aap}, \href{https://ui.adsabs.harvard.edu/abs/2012A&A...537A.146E}{537, A146}

\bibitem[{{Fender} {et~al.}(2004){Fender}, {Belloni}, \& {Gallo}}]{Fender2004}
{Fender}, R.~P., {Belloni}, T.~M., \& {Gallo}, E. 2004, \href{http://dx.doi.org/10.1111/j.1365-2966.2004.08384.x}{\color{magenta}\mnras}, \href{https://ui.adsabs.harvard.edu/abs/2004MNRAS.355.1105F}{355, 1105}

\bibitem[{{Fortin} {et~al.}(2022){Fortin}, {Garc{\'\i}a}, \& {Chaty}}]{fortin2022}
{Fortin}, F., {Garc{\'\i}a}, F., \& {Chaty}, S. 2022, \href{http://dx.doi.org/10.1051/0004-6361/202244048}{\color{magenta}\aap}, \href{https://ui.adsabs.harvard.edu/abs/2022A&A...665A..69F}{665, A69}

\bibitem[{{Gaia Collaboration} {et~al.}(2021){Gaia Collaboration}, {Brown}, {Vallenari}, {Prusti}, {de Bruijne}, {Babusiaux}, {Biermann}, {Creevey}, {Evans}, {Eyer}, {Hutton}, {Jansen}, {Jordi}, {Klioner}, {Lammers}, {Lindegren}, {Luri}, {Mignard}, {Panem}, {Pourbaix}, {Randich}, {Sartoretti}, {Soubiran}, {Walton}, {Arenou}, {Bailer-Jones}, {Bastian}, {Cropper}, {Drimmel}, {Katz}, {Lattanzi}, {van Leeuwen}, {Bakker}, {Cacciari}, {Casta{\~n}eda}, {De Angeli}, {Ducourant}, {Fabricius}, {Fouesneau}, {Fr{\'e}mat}, {Guerra}, {Guerrier}, {Guiraud}, {Jean-Antoine Piccolo}, {Masana}, {Messineo}, {Mowlavi}, {Nicolas}, {Nienartowicz}, {Pailler}, {Panuzzo}, {Riclet}, {Roux}, {Seabroke}, {Sordo}, {Tanga}, {Th{\'e}venin}, {Gracia-Abril}, {Portell}, {Teyssier}, {Altmann}, {Andrae}, {Bellas-Velidis}, {Benson}, {Berthier}, {Blomme}, {Brugaletta}, {Burgess}, {Busso}, {Carry}, {Cellino}, {Cheek}, {Clementini}, {Damerdji}, {Davidson}, {Delchambre}, {Dell'Oro}, {Fern{\'a}ndez-Hern{\'a}ndez}, {Galluccio}, {Garc{\'\i}a-Lario},
  {Garcia-Reinaldos}, {Gonz{\'a}lez-N{\'u}{\~n}ez}, {Gosset}, {Haigron}, {Halbwachs}, {Hambly}, {Harrison}, {Hatzidimitriou}, {Heiter}, {Hern{\'a}ndez}, {Hestroffer}, {Hodgkin}, {Holl}, {Jan{\ss}en}, {Jevardat de Fombelle}, {Jordan}, {Krone-Martins}, {Lanzafame}, {L{\"o}ffler}, {Lorca}, {Manteiga}, {Marchal}, {Marrese}, {Moitinho}, {Mora}, {Muinonen}, {Osborne}, {Pancino}, {Pauwels}, {Petit}, {Recio-Blanco}, {Richards}, {Riello}, {Rimoldini}, {Robin}, {Roegiers}, {Rybizki}, {Sarro}, {Siopis}, {Smith}, {Sozzetti}, {Ulla}, {Utrilla}, {van Leeuwen}, {van Reeven}, {Abbas}, {Abreu Aramburu}, {Accart}, {Aerts}, {Aguado}, {Ajaj}, {Altavilla}, {{\'A}lvarez}, {{\'A}lvarez Cid-Fuentes}, {Alves}, {Anderson}, {Anglada Varela}, {Antoja}, {Audard}, {Baines}, {Baker}, {Balaguer-N{\'u}{\~n}ez}, {Balbinot}, {Balog}, {Barache}, {Barbato}, {Barros}, {Barstow}, {Bartolom{\'e}}, {Bassilana}, {Bauchet}, {Baudesson-Stella}, {Becciani}, {Bellazzini}, {Bernet}, {Bertone}, {Bianchi}, {Blanco-Cuaresma}, {Boch}, {Bombrun}, {Bossini},
  {Bouquillon}, {Bragaglia}, {Bramante}, {Breedt}, {Bressan}, {Brouillet}, {Bucciarelli}, {Burlacu}, {Busonero}, {Butkevich}, {Buzzi}, {Caffau}, {Cancelliere}, {C{\'a}novas}, {Cantat-Gaudin}, {Carballo}, {Carlucci}, {Carnerero}, {Carrasco}, {Casamiquela}, {Castellani}, {Castro-Ginard}, {Castro Sampol}, {Chaoul}, {Charlot}, {Chemin}, {Chiavassa}, {Cioni}, {Comoretto}, {Cooper}, {Cornez}, {Cowell}, {Crifo}, {Crosta}, {Crowley}, {Dafonte}, {Dapergolas}, {David}, \& {David}}]{gaia2021}
{Gaia Collaboration}, {Brown}, A.~G.~A., {Vallenari}, A., {et~al.} 2021, \href{http://dx.doi.org/10.1051/0004-6361/202039657}{\color{magenta}\aap}, \href{https://ui.adsabs.harvard.edu/abs/2021A&A...649A...1G}{649, A1}

\bibitem[{{Gallo} {et~al.}(2006){Gallo}, {Fender}, {Miller-Jones}, {Merloni}, {Jonker}, {Heinz}, {Maccarone}, \& {van der Klis}}]{gallo2006}
{Gallo}, E., {Fender}, R.~P., {Miller-Jones}, J.~C.~A., {et~al.} 2006, \href{http://dx.doi.org/10.1111/j.1365-2966.2006.10560.x}{\color{magenta}\mnras}, \href{https://ui.adsabs.harvard.edu/abs/2006MNRAS.370.1351G}{370, 1351}

\bibitem[{{Guedel} \& {Benz}(1993)}]{gb1993}
{Guedel}, M. \& {Benz}, A.~O. 1993, \href{http://dx.doi.org/10.1086/186766}{\color{magenta}\apjl}, \href{https://ui.adsabs.harvard.edu/abs/1993ApJ...405L..63G}{405, L63}

\bibitem[{{Hoogerwerf} {et~al.}(2001){Hoogerwerf}, {de Bruijne}, \& {de Zeeuw}}]{hoogerwerf2001}
{Hoogerwerf}, R., {de Bruijne}, J.~H.~J., \& {de Zeeuw}, P.~T. 2001, \href{http://dx.doi.org/10.1051/0004-6361:20000014}{\color{magenta}\aap}, \href{https://ui.adsabs.harvard.edu/abs/2001A&A...365...49H}{365, 49}

\bibitem[{{Janssens} {et~al.}(2023){Janssens}, {Shenar}, {Degenaar}, {Bodensteiner}, {Sana}, {Audenaert}, \& {Frost}}]{Janssens2023}
{Janssens}, S., {Shenar}, T., {Degenaar}, N., {et~al.} 2023, \href{http://dx.doi.org/10.1051/0004-6361/202347318}{\color{magenta}\aap}, \href{https://ui.adsabs.harvard.edu/abs/2023A&A...677L...9J}{677, L9}

\bibitem[{{Lucarelli} {et~al.}(2010){Lucarelli}, {Verrecchia}, {Striani}, {Pittori}, {Tavani}, {Vercellone}, {Bulgarelli}, {Gianotti}, {Trifoglio}, {Chen}, {Giuliani}, {Mereghetti}, {Caraveo}, {Perotti}, {Donnarumma}, {D'Ammando}, {Del Monte}, {Evangelista}, {Feroci}, {Lazzarotto}, {Pacciani}, {Soffitta}, {Costa}, {Lapshov}, {Rapisarda}, {Argan}, {Piano}, {Pucella}, {Sabatini}, {Trois}, {Vittorini}, {Fuschino}, {Galli}, {Labanti}, {Marisaldi}, {Di Cocco}, {Pellizzoni}, {Pilia}, {Barbiellini}, {Longo}, {Moretti}, {Vallazza}, {Morselli}, {Picozza}, {Prest}, {Lipari}, {Zanello}, {Cattaneo}, {Rappoldi}, {Santolamazza}, {Colafrancesco}, {Giommi}, \& {Salotti}}]{Lucarelli2010}
{Lucarelli}, F., {Verrecchia}, F., {Striani}, E., {et~al.} 2010, The Astronomer's Telegram, \href{https://ui.adsabs.harvard.edu/abs/2010ATel.2761....1L}{2761, 1}

\bibitem[{{Massi} \& {Kaufman Bernad{\'o}}(2008)}]{Massi2008}
{Massi}, M. \& {Kaufman Bernad{\'o}}, M. 2008, \href{http://dx.doi.org/10.1051/0004-6361:20077567}{\color{magenta}\aap}, \href{https://ui.adsabs.harvard.edu/abs/2008A&A...477....1M}{477, 1}

\bibitem[{{Merloni} {et~al.}(2003){Merloni}, {Heinz}, \& {di Matteo}}]{Merloni2003}
{Merloni}, A., {Heinz}, S., \& {di Matteo}, T. 2003, \href{http://dx.doi.org/10.1046/j.1365-2966.2003.07017.x}{\color{magenta}\mnras}, \href{https://ui.adsabs.harvard.edu/abs/2003MNRAS.345.1057M}{345, 1057}

\bibitem[{{Mirabel}(2007)}]{mirabel2007}
{Mirabel}, I.~F. 2007, \href{http://dx.doi.org/10.1007/s10509-007-9459-y}{\color{magenta}\apss}, \href{https://ui.adsabs.harvard.edu/abs/2007Ap&SS.309..267M}{309, 267}

\bibitem[{{Moffat} {et~al.}(1998){Moffat}, {Marchenko}, {Seggewiss}, {van der Hucht}, {Schrijver}, {Stenholm}, {Lundstrom}, {Setia Gunawan}, {Sutantyo}, {van den Heuvel}, {de Cuyper}, \& {Gomez}}]{moffat1998}
{Moffat}, A.~F.~J., {Marchenko}, S.~V., {Seggewiss}, W., {et~al.} 1998, \aap, \href{https://ui.adsabs.harvard.edu/abs/1998A&A...331..949M}{331, 949}

\bibitem[{{Mold{\'o}n}(2012)}]{Moldon2012}
{Mold{\'o}n}, F.~J. 2012, \href{https://ui.adsabs.harvard.edu/abs/2012PhDT.......535M}{{Structure and nature of gamma-ray binaries by means of VLBI observations}}, PhD thesis, University of Barcelona, Spain

\bibitem[{{Munar-Adrover} {et~al.}(2016){Munar-Adrover}, {Sabatini}, {Piano}, {Tavani}, {Nguyen}, {Lucarelli}, {Verrecchia}, \& {Pittori}}]{Munar-Adrover2016}
{Munar-Adrover}, P., {Sabatini}, S., {Piano}, G., {et~al.} 2016, \href{http://dx.doi.org/10.3847/0004-637X/829/2/101}{\color{magenta}\apj}, \href{https://ui.adsabs.harvard.edu/abs/2016ApJ...829..101M}{829, 101}

\bibitem[{{Plotkin} {et~al.}(2017){Plotkin}, {Miller-Jones}, {Gallo}, {Jonker}, {Homan}, {Tomsick}, {Kaaret}, {Russell}, {Heinz}, {Hodges-Kluck}, {Markoff}, {Sivakoff}, {Altamirano}, \& {Neilsen}}]{plotkin2017}
{Plotkin}, R.~M., {Miller-Jones}, J.~C.~A., {Gallo}, E., {et~al.} 2017, \href{http://dx.doi.org/10.3847/1538-4357/834/2/104}{\color{magenta}\apj}, \href{https://ui.adsabs.harvard.edu/abs/2017ApJ...834..104P}{834, 104}

\bibitem[{{Reid} {et~al.}(2009){Reid}, {Menten}, {Zheng}, {Brunthaler}, {Moscadelli}, {Xu}, {Zhang}, {Sato}, {Honma}, {Hirota}, {Hachisuka}, {Choi}, {Moellenbrock}, \& {Bartkiewicz}}]{reid2009}
{Reid}, M.~J., {Menten}, K.~M., {Zheng}, X.~W., {et~al.} 2009, \href{http://dx.doi.org/10.1088/0004-637X/700/1/137}{\color{magenta}\apj}, \href{https://ui.adsabs.harvard.edu/abs/2009ApJ...700..137R}{700, 137}

\bibitem[{{Rib{\'o}} {et~al.}(2017){Rib{\'o}}, {Munar-Adrover}, {Paredes}, {Marcote}, {Iwasawa}, {Mold{\'o}n}, {Casares}, {Migliari}, \& {Paredes-Fortuny}}]{Ribo2017}
{Rib{\'o}}, M., {Munar-Adrover}, P., {Paredes}, J.~M., {et~al.} 2017, \href{http://dx.doi.org/10.3847/2041-8213/835/2/L33}{\color{magenta}\apjl}, \href{https://ui.adsabs.harvard.edu/abs/2017ApJ...835L..33R}{835, L33}, [R17]

\bibitem[{{Rivinius} {et~al.}(2024){Rivinius}, {Klement}, {Chojnowski}, {Baade}, {Shepard}, \& {Hadrava}}]{Rivinius2024}
{Rivinius}, T., {Klement}, R., {Chojnowski}, S.~D., {et~al.} 2024, in IAU Symposium, Vol. 361, Massive Stars Near and Far, ed. {Mackey}, J., {Vink}, J.~S., \& {St-Louis}, N., \href{https://ui.adsabs.harvard.edu/abs/2024IAUS..361..332R}{332--333}

\bibitem[{{Sch{\"o}nrich} {et~al.}(2010){Sch{\"o}nrich}, {Binney}, \& {Dehnen}}]{schonrich2010}
{Sch{\"o}nrich}, R., {Binney}, J., \& {Dehnen}, W. 2010, \href{http://dx.doi.org/10.1111/j.1365-2966.2010.16253.x}{\color{magenta}\mnras}, \href{https://ui.adsabs.harvard.edu/abs/2010MNRAS.403.1829S}{403, 1829}

\bibitem[{{van den Eijnden} {et~al.}(2021){van den Eijnden}, {Degenaar}, {Russell}, {Wijnands}, {Bahramian}, {Miller-Jones}, {Hern{\'a}ndez Santisteban}, {Gallo}, {Atri}, {Plotkin}, {Maccarone}, {Sivakoff}, {Miller}, {Reynolds}, {Russell}, {Maitra}, {Heinke}, {Armas Padilla}, \& {Shaw}}]{Eijnden2021}
{van den Eijnden}, J., {Degenaar}, N., {Russell}, T.~D., {et~al.} 2021, \href{http://dx.doi.org/10.1093/mnras/stab1995}{\color{magenta}\mnras}, \href{https://ui.adsabs.harvard.edu/abs/2021MNRAS.507.3899V}{507, 3899}

\bibitem[{{Williams} {et~al.}(2010){Williams}, {Gies}, {Matson}, {Touhami}, {Grundstrom}, {Huang}, \& {McSwain}}]{Williams2010}
{Williams}, S.~J., {Gies}, D.~R., {Matson}, R.~A., {et~al.} 2010, \href{http://dx.doi.org/10.1088/2041-8205/723/1/L93}{\color{magenta}\apjl}, \href{https://ui.adsabs.harvard.edu/abs/2010ApJ...723L..93W}{723, L93}

\bibitem[{{Yanza} {et~al.}(2022){Yanza}, {Masqu{\'e}}, {Dzib}, {Rodr{\'\i}guez}, {Medina}, {Kurtz}, {Loinard}, {Trinidad}, {Menten}, \& {Rodr{\'\i}guez-Rico}}]{yanza2022}
{Yanza}, V., {Masqu{\'e}}, J.~M., {Dzib}, S.~A., {et~al.} 2022, \href{http://dx.doi.org/10.3847/1538-3881/ac67ec}{\color{magenta}\aj}, \href{https://ui.adsabs.harvard.edu/abs/2022AJ....163..276Y}{163, 276}

\end{thebibliography}

\end{document}